\documentstyle[12pt,epsf]{article}
\begin{document}
\topskip -1mm
\title{\bf Monte Carlo Algorithms based on the Number of Potential Moves}

\author{Jian-Sheng Wang\thanks{email: wangjs@cz3.nus.edu.sg} 
         \ and Lik Wee Lee\\  
Department of Computational Science,\\
National University of Singapore,\\
Singapore 119260, Republic of Singapore}

\date{12 March 1999}

\maketitle

\begin{abstract}
We discuss Monte Carlo dynamics based on $\langle N(\sigma,
\Delta E) \rangle_E$, the (microcanonical) average number of
potential moves which increase the energy by $\Delta E$ in a
single spin flip.  The microcanonical average can be sampled
using Monte Carlo dynamics of a single spin flip with a
transition rate $\min(1, \langle N(\sigma', E-E')\rangle_{E'} /
\langle N(\sigma, E'-E) \rangle_E)$ from energy $E$ to $E'$.  A
cumulative average (over Monte Carlo steps) can be used as a
first approximation to the exact microcanonical average in the
flip rate.  The associated histogram is a constant independent of
the energy.  The canonical distribution of energy can be obtained
from the transition matrix Monte Carlo dynamics.  This second
dynamics has fast relaxation time---at the critical temperature
the relaxation time is proportional to specific heat.  The
dynamics are useful in connection with reweighting methods for
computing thermodynamic quantities.


\end{abstract}

\section{Introduction}

The traditional Monte Carlo method applied to statistical physics
\cite{Kalos,Binder} is mostly a sampling method to generate standard 
statistical ensembles, e.g., the canonical ensemble or
microcanonical ensemble.  In recent years, other ensembles have
been used which do not correspond to thermodynamically meaningful
ensembles, but used only as a vehicle for computing quantities of
interests by Monte Carlo method.  The earliest such method is the
umbrella sampling \cite{umbrella}.  Other important recent
developments are the multi-canonical simulations \cite{Berg,Lee},
$1/k$-sampling \cite{stinchcombe}, and broad histogram methods
\cite{oliveira}.  According to one definition \cite{berg-def},
the multi-canonical ensemble is an ensemble that the probability
$P(E)$ of having energy $E$ is a constant.  This may be realized
in a piecewise fashion.  In the $1/k$-sampling method, the
probability for a state having energy $E$ is given by $1/
\sum_{E'<E} n(E')$, where $n(E)$ is density of states.  The broad
histogram dynamics does not have a well characterized
distribution, but $P(E)$ is much broader than the canonical
ensemble.  The canonical distribution is well approximated by a
Gaussian function.  It was also pointed out that this dynamics is not
entirely correct \cite{berg-hansmann,wang-EPJB}.  The ultimate
goal of generating these distributions is usually to compute
thermodynamic averages, which are mostly canonical averages.
Some form of reweighting is then used to obtain the desired
distribution.

Recently, we proposed a dynamics \cite{wang-EPJB} which can
generate a flat histogram $P(E) = const$.  This method is exact
when a ``self-consistency'' is achieved.  The meaning of which
will be made clear later. Similar to the broad histogram method,
the central quantity is $\langle N(\sigma,\Delta E)\rangle_E$ ,
the microcanonical average of the number of ways to move from one
energy level $E$, to a nearby energy level, $E + \Delta E$.  We
can then construct either the density of states or the canonical
distribution at any temperature.  The canonical distribution is
determined from an artificial dynamics which we call it
transition matrix Monte Carlo.  We discuss these methods and
present some preliminary results in the later section.

\section{Sampling the Inverse Density of States}

We illustrate the method using a two-dimensional Ising model on a
square lattice as an example.  First of all, we choose a type of
permissible moves.  For purpose of connection with standard
single-spin-flip Glauber (or Metropolis) dynamics, we take the
set of moves to be all single-spin flips.  For a given state
$\sigma$, we can obtain $N=L^2$ new states through flipping each
of the spins in an $L \times L$ system.  If the original state
has energy $E = E(\sigma)$, the new state may have energy $E +
\Delta E$.  Since the energy spectrum is discrete, we have only a
finite number of possibilities for the new energies; for the
two-dimensional Ising model, we have five possible energy
changes, $\Delta E = 0, \pm 4 J, \pm 8J$. Let the counts of moves
for each energy changes be $N(\sigma, \Delta E)$. Hence the total
number of moves is $\sum_{\Delta E} N(\sigma, \Delta E) = N$.

Following the argument of Oliveira \cite{oliveira}, we consider two
energy levels $E$ and $E' = E + \Delta E$. Each move from the
state $\sigma$ of energy $E$ to the state $\sigma'$ of energy
$E'$ is through a single spin flip and the reverse move is also
allowed.  Thus, the total number of moves from all the states
with energy $E$ to $E'$ is the same as from $E'$ to $E$:
\begin{equation}
  \sum_{E(\sigma) = E} N(\sigma, \Delta E)
= \sum_{E(\sigma') = E + \Delta E} N(\sigma', -\Delta E).
\label{eq:reversible}
\end{equation}
The microcanonical average of a quantity $A(\sigma)$ is defined
as
\begin{equation}
  \langle A \rangle_E =  {1 \over n(E) } \sum_{E(\sigma) = E} A(\sigma),
\end{equation}
where the summation is over all the states with a fixed energy
$E$, and $n(E)$ is the number of such states.  In terms of
microcanonical average, we can re-write Eq.~(\ref{eq:reversible})
as
\begin{equation}
  n(E) \langle N(\sigma, \Delta E) \rangle_E
= n(E + \Delta E) \langle N(\sigma', -\Delta E) \rangle_{E+\Delta E}.
\label{eq:broad}
\end{equation}
This is the basic equation of the broad histogram method
\cite{oliveira} and is also our starting point for a flat
histogram sampling algorithm.

Consider the following flip rate for a single-spin-flip move from
state $\sigma$ to $\sigma'$ with energy $E$ and $E'= E + \Delta
E$, respectively:
\begin{equation}
r(E'|E) = \min\left(1, {\langle N(\sigma', -\Delta E) \rangle_{E'} \over
                   \langle N(\sigma, \Delta E) \rangle_E } \right).
\end{equation}
The site of the spin flip is chosen at random.  Then the detailed
balance condition for this rate
\begin{equation}
  r(E'|E) P(\sigma) = r(E|E') P(\sigma') 
\end{equation}  
is satisfied for $P(\sigma) \propto 1/n(E(\sigma))$.  Thus energy
histogram is flat, 
\begin{equation}
P(E) = \sum_{ E(\sigma) = E} P(\sigma) \propto
 n(E)  {1 \over n(E) } = const.
\end{equation}

Suppose that such samples are generated, then in some sense, it
is the optimal ensemble for evaluation of $\langle N(\sigma, \Delta
E) \rangle_E$.  This is because for different $E$, we take
samples uniformly in $E$, and thus the relative errors in
$\langle N(\sigma, \Delta E) \rangle_E$ are about the same for
all $E$.

Since $\langle N(\sigma, \Delta E) \rangle_E$ is not known in
general, we cannot start the simulation unless an approximation
scheme is used.  We can think of the process as finding fixed
point value of the system $x = f(x)$, where vector $x$ represents
the whole set of $\langle N(\sigma, \Delta E) \rangle_E$ values.
While the function $f$ can be evaluated, its explicit form is not
known.  Some iterative scheme may be useful to speed-up the
convergence.  To start the iterative process, we use a cumulative
average for the true microcanonical average.  For those $E$ which
we do not have any sample yet, we simply set $r(E'|E)$ to 1.
This simple scheme is very good for small systems even without
iteration.

\section{The transition matrix Monte Carlo dynamics}

We can construct a Monte Carlo dynamics, in the space of energy,
with the average number of moves, $\langle N(\sigma, \Delta E)
\rangle_E$, \cite{Wang-Tay-Swendsen}.  Let us look at a 
single-spin-flip Glauber dynamics. Suppose we do not care about
the spin states and only want to know the change of energy.  The
rate of a spin flip is given by the Glauber rate,
\begin{equation}
w(\Delta E) =  {1\over 2}\left[ 1 - \tanh{ \Delta E \over 2kT} \right].
\label{eq:glauber}
\end{equation}
Since there are (on average) $\langle N(\sigma, \Delta E)
\rangle_E$ different ways of going from $E$ to $E'= E + \Delta
E$, the total probability for transition from $E$ to $E'$ is
\begin{equation}
 W(E'|E) = w(\Delta E) \langle N(\sigma, \Delta E) \rangle_E, 
\quad E \neq E'.
\end{equation}
The diagonal elements are fixed by the requirement that $W(E|E')$
is a stochastic matrix.  This transition matrix satisfies
detailed balance with respect to the canonical distribution,
$P_T(E) \propto n(E) \exp(-E/kT)$. Thus the stationary
distribution is the canonical distribution.  This new dynamics in
the space of energy $E$ is related to the single-spin-flip
dynamics by
\cite{Wang-Tay-Swendsen}
\begin{equation}
W(E'|E) = {1 \over n(E) } \sum_{E(\sigma) = E}\>\sum_{E(\sigma') = E'}
\!\!\! \Gamma(\sigma'| \sigma),
\end{equation} 
where $\Gamma(\sigma'| \sigma)$ is the transition matrix of the
single-spin-flip dynamics.

\begin{figure}[tb]
\center{\leavevmode\epsfxsize=0.75\hsize\epsfbox{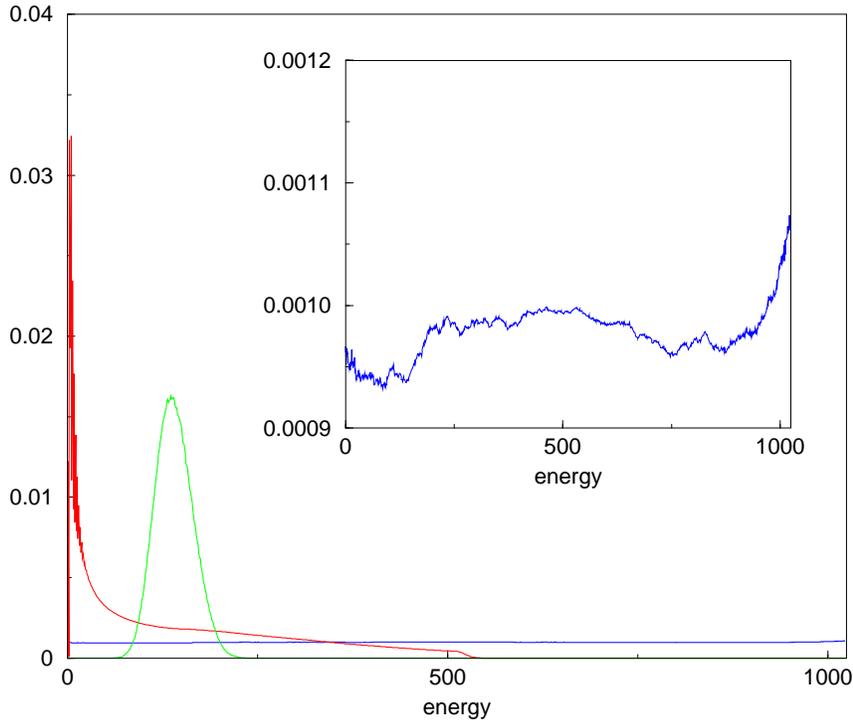}}
\caption[histogram]{Three different types of histograms: (a) canonical
simulations, (b) broad histogram, (c) flat histogram, on a $32
\times 32$ lattice and $10^7$ Monte Carlo steps each.  The insert
shows flat histogram on a finer scale.  The mean is $9.8 \times
10^{-4}$ with standard deviation $2.1 \times 10^{-5}$.}
\label{fig:histograms}
\end{figure}

An interesting aspect of this dynamics is that it has a much reduced
critical slowing down.  In fact, one can show that the relaxation
time at the critical point $T_c$ is proportional to the specific
heat. Thus for the two-dimensional Ising model, the divergence of
the relaxation time is only logarithmic.  In one dimension, the
dynamics has a curious dynamical critical exponent of $z=1$ as
oppose to 2 for the local dynamics and 0 for the Swendsen-Wang
dynamics \cite{SW}.  Since the dynamics can not be realized
without first knowing the values $\langle N(\sigma, \Delta E)
\rangle_E$, the real usefulness is in the construction of
canonical distribution from the samples obtained by flat
histogram or any other algorithms that can compute $\langle
N(\sigma, \Delta E) \rangle_E$ accurately.

\section{Results}

In Fig.~\ref{fig:histograms}, we show the energy histograms for
three different types of dynamics of the $32 \times 32$
two-dimensional Ising model, (a) the Gaussian-like peak for the
standard canonical ensemble at the critical temperature $T_c$;
(b) the broad histogram dynamics with a sharp peak near $E=0$;
(c) the flat histogram method with an insert showing the
fluctuation on a fine scale.

Given the estimates for $\langle N(\sigma, \Delta E) \rangle_E$,
there are a number of ways to determine the canonical
distribution, $P_T(E) \propto n(E) \exp(-E/kT)$.  For example, we
can use Eq.~(\ref{eq:broad}) to determine the density of states.
We can also determine $P_T(E)$ directly from the detailed balance
of the transition matrix Monte Carlo dynamics
\cite{Wang-Tay-Swendsen}:
\begin{equation}
   w(\Delta E) \langle N(\sigma, \Delta E) \rangle_E  P_T(E) =
w(-\Delta E) \langle N(\sigma', -\Delta E) \rangle_{E+\Delta E}  
P_T(E+\Delta E),
\end{equation}
where $w(\Delta E)$ is given by Eq.~(\ref{eq:glauber}).  Since
there are more equations than unknowns, it is natural to solve
these over-determined equations with least-square method.
However, a more direct iterative scheme is also quite accurate
and more efficient.

\begin{figure}[tb]
\center{\leavevmode\epsfxsize=0.9\hsize\epsfbox{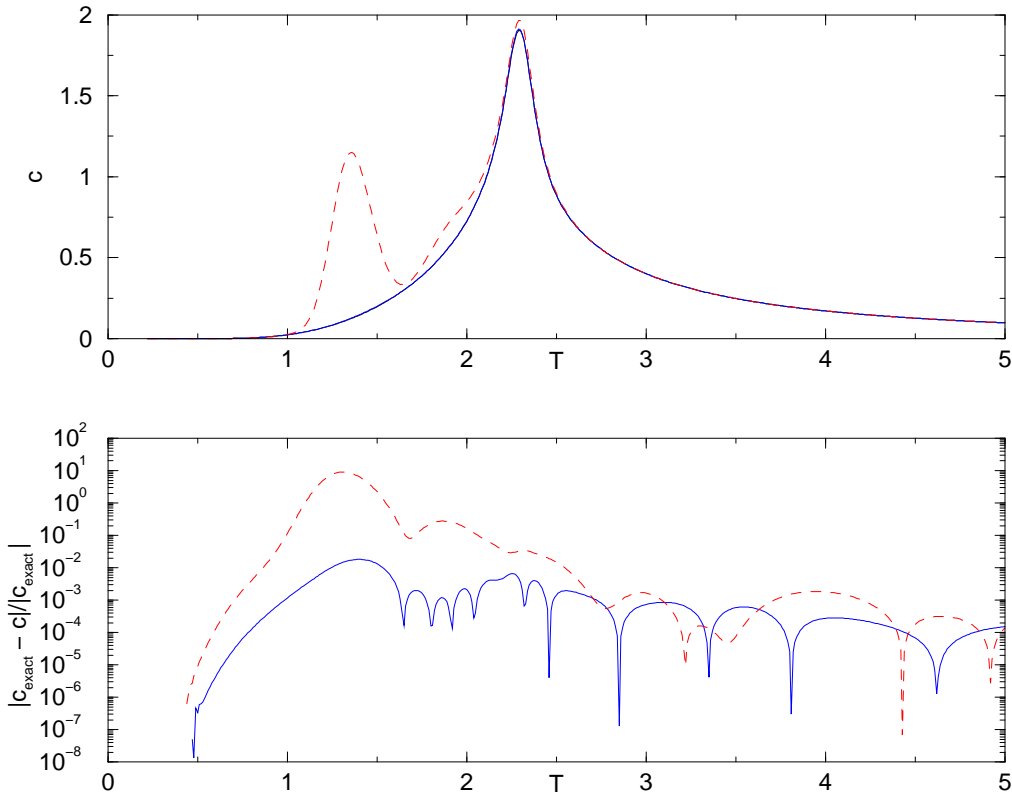}}
\caption[specific heat]{Specific heat of a $32 \times 32$ Ising Model.
$10^7$ Monte Carlo steps were used in both the broad histogram
(dashed lines) and the flat histogram (solid lines) method.}
\label{fig:heat}
\end{figure}

In Fig.~\ref{fig:heat}, we show the specific heat (upper part)
and relative errors as compared with exact results \cite{Beale}.
The dash lines are for the broad histogram method and solid lines
are from the flat histogram sampling.  The broad histogram method
shows an anomalous peak around $T=1.3$, while the flat histogram
result agrees with exact values with errors of $10^{-2}$ or less.

\begin{figure}[tb]
\center{\leavevmode\epsfxsize=0.9\hsize\epsfbox{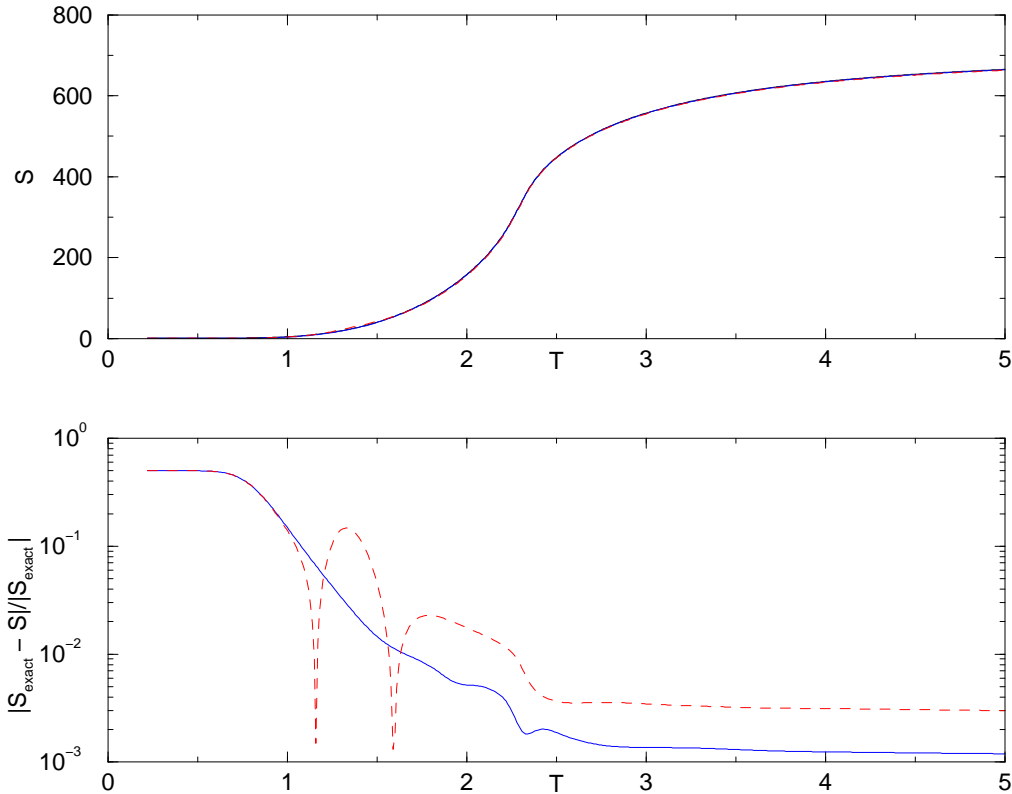}}
\caption[entropy]{Entropy of a $32 \times 32$ Ising Model. $10^7$ Monte
Carlo steps were used in the broad histogram (dashed lines) and flat
histogram (solid lines) method.}
\label{fig:S}
\end{figure}

Since we can compute the density of states $n(E)$ easily, we can
also compute free energy and entropy with ease.  These quantities
are more difficult to compute by the traditional methods.
Fig.~\ref{fig:S} shows the entropy and errors.  The flat
histogram is again better than the broad histogram method.

\begin{figure}[tb]
\center{\leavevmode\epsfxsize=0.75\hsize\epsfbox{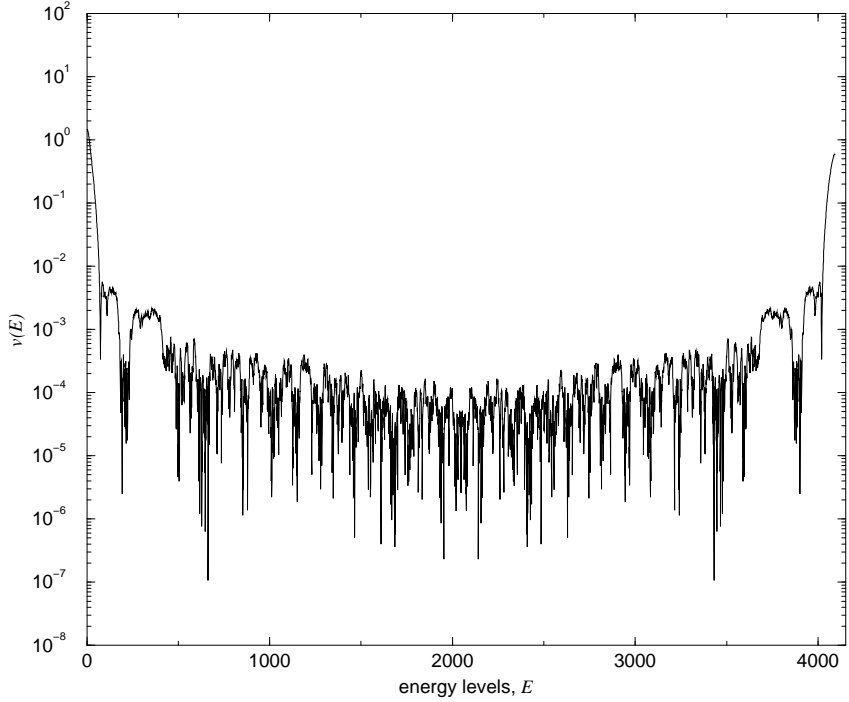}}
\caption[violation]{Detailed balance violation of $64 \times 64$ 
Ising Model.  $10^8$ Monte Carlo steps were used in flat histogram
method.}
\label{fig:v}
\end{figure}

All approaches that use reweighting technique, such as the
histogram methods of Ferrenberg and Swendsen
\cite{Ferrenberg-Swendsen}, Lee's version of multicanonical
method \cite{Lee}, or the broad histogram method \cite{oliveira},
have the problem of scalability for large systems.  Our flat
histogram method also suffers from this.  While the simple method
without an iterative process and without requirement for
self-consistency seems to work well for systems $L \leq 32$,
systematic errors are observed for large systems.  Substantial
deviations (extra anomalous peaks in the specific heat, for
example) are present for the $L=64$ systems.  Such systematic
deviations can be measured quantitatively by what we called
detailed balance violation \cite{wang-EPJB}:
\begin{equation}
     v(E) = \left| 1 - { g(E,E'') g(E'', E') g(E', E) \over
                    g(E,E') g(E', E'') g(E'', E) } \right|
\end{equation}
where $g$ is generally a transition rate---for our problem here,
we'll take $g(E, E') = \langle N(\sigma, E'-E) \rangle_E$, with
$E' = E + 4J$ and $E''= E + 8J$.  The quantity $v(E)$ should be
zero, up to the usual Monte Carlo statistical errors, if the
estimates are not systematically biased.  In Fig.~\ref{fig:v}, we
show this quantity as a function of $E$ for the $L=64$ system.
The largest violation occurs at the two ends of the distribution.
This systematic trend is also present for small systems.

There are a number of ways to fix this problem. One is to do a
number of canonical simulations at lower temperatures where the
violation of detailed balance is biggest.  This indeed proves to
be effective for the Ising model.  However, this solution is not
very satisfactory, as such simulations may be very difficult, for
example, for spin glasses.  Thus, a more systematic approach is
to use an iterative scheme, which can hopefully converge to the
true value without any systematic bias.

\section{Conclusion}

We study a recently proposed Monte Carlo dynamics in which the
energy histogram is exactly flat in principle.  We demonstrated
that such method is capable of giving highly accurate results for
the thermodynamic quantities in a single or few simulations for
the whole temperature region. While some systematic errors are
present in our current simple implementation, there are ways to
improve the naive algorithm. We expect that this method will be a
useful alternative for thermodynamic calculations, especially for
free energy and entropy calculations.

\end{document}